# 自律分散型センサフュージョンのためのバイアス推定

## マルチエージェント型バイアス推定法

古川　英俊[†]

† 株式会社東芝 インフラシステムソリューション社
〒 212–8581 川崎市幸区小向東芝町 1 番地
E-mail: †hidetoshi.furukawa@toshiba.co.jp

**あらまし**　目標を観測する複数センサがネットワークに接続されたセンサネットワークにおいてセンサフュージョンを行う場合，センサのバイアスが追跡している目標の相関処理や統合処理の精度に影響を与える．このため，センサのバイアスを推定・補償する方法として，センサからの観測値を一箇所に集めて，カルマンフィルタによりセンサフュージョンとバイアス推定を同時に実施する方法が提案されている．一方，センサネットワークにおけるセンサフュージョンの構成は，センサからの観測値を統合する集中型や，各センサの追跡フィルタで処理した航跡を統合する分散型のほか，集中管理機構を必要としない自律分散型も提案されている．本稿では，自律分散型センサフュージョンにおいてバイアスを推定・補償するためのマルチエージェント型バイアス推定法を提案する．
**キーワード**　センサネットワーク，センサフュージョン，バイアス，マルチエージェント

# Bias Estimation for Decentralized Sensor Fusion

## Multi-Agent Based Bias Estimation Method

Hidetoshi FURUKAWA[†]

† Toshiba Corporation Infrastructure Systems & Solutions Company
1 Komukaitoshiba-cho, Saiwai-ku Kawasaki-shi, Kanagawa 212–8581, Japan
E-mail: †hidetoshi.furukawa@toshiba.co.jp

**Abstract**　In multi-sensor data fusion (or sensor fusion), sensor biases (or offsets) often affect the accuracy of the correlation and integration results of the tracking targets. Therefore, to estimate and compensate the bias, several methods are proposed. However, most methods involve bias estimation and sensor fusion simultaneously by using Kalman filter after collecting the plot data together. Hence, these methods cannot support to fuse the track data prepared by tracking filter at each sensor node. This report proposes the new bias estimation method based on multi-agent model, in order to estimate and compensate the bias for decentralized sensor fusion.
**Key words**　sensor network, sensor fusion, bias, multi-agent

## 1. はじめに

目標を観測する複数センサがネットワークに接続されたセンサネットワーク（sensor network）においてセンサフュージョン（sensor fusion）を行う場合，センサのバイアス（bias）が追跡している目標の相関処理や統合処理の精度に影響を与える．このため，センサのバイアスを推定・補償する方法として，文献 [1] には，センサからの観測値（plot）を一箇所に集めて，カルマンフィルタによりセンサフュージョンとバイアス推定を同時に実施する方法（以降，従来法と呼ぶ）が提案されている．

一方，センサネットワークにおけるセンサフュージョンの構成は，図 1 に示すように，センサからの観測値を一箇所に集めて統合する集中型（centralized）のセンサフュージョンである観測値融合（plot fusion）や，各センサの追跡フィルタで処理した航跡（track）を一箇所に集めて統合する分散型（distributed）のセンサフュージョンである航跡融合（track fusion）のほか，集中管理機構を必要としない自律分散型（decentralized）のセンサフュージョンである航跡融合も示されている [2]．従来法は，観測値融合を行う集中型センサフュージョン（centralized sensor fusion）に適用することはできるが，航跡融合を行う分

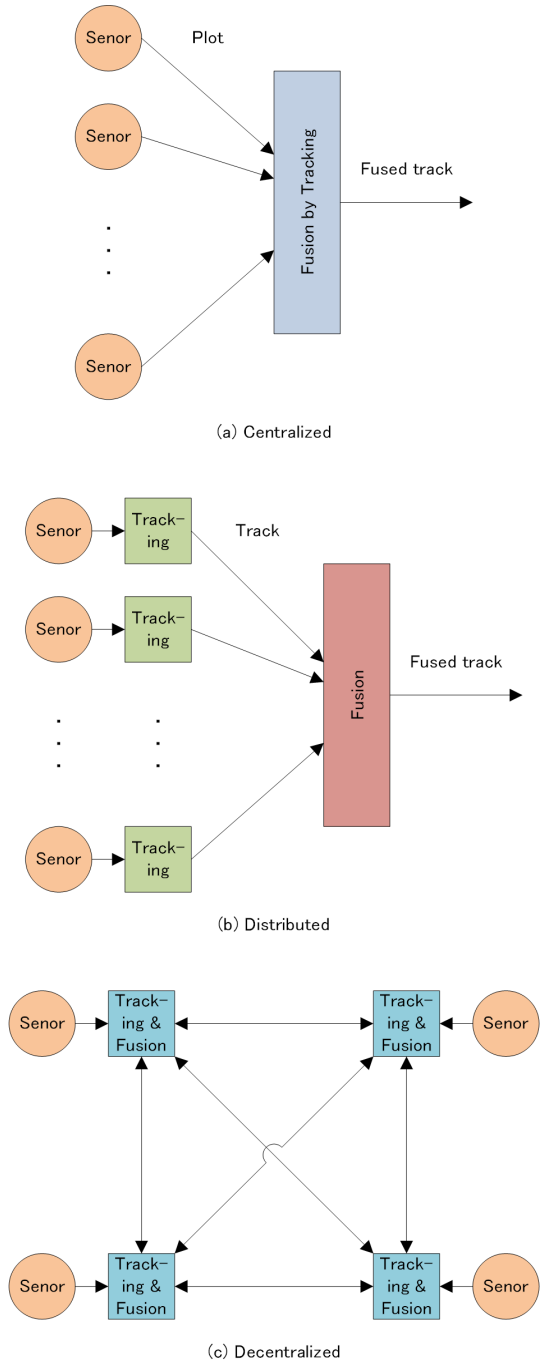

図 1 センサフュージョンの構成の分類
Fig. 1 Structures of sensor fusion.

散型センサフュージョン（distributed sensor fusion）や自律分散型センサフュージョン（decentralized sensor fusion）に適用することはできない．

本稿では，自律分散型センサフュージョンにおいてバイアスを推定・補償するためのマルチエージェント型バイアス推定法を提案する．

## 2. マルチエージェント型バイアス推定法

マルチエージェント型バイアス推定法とは，センサネットワーク全体を，複数エージェントの局所的な相互作用をもとに大域的な機能を発現するマルチエージェントシステム [3] とし

て捉え，$N$ 個のセンサ $S_i$ ($i \in \{1, ..., N\}$) から構成されるセンサネットワークにおいて，センサ $S_i$ に対応するエージェント $A_i$ が，グローバル座標系の目標状態ベクトルを，他センサ $S_j$ ($j \in \{1, ..., N\}, j \neq i$) に対応するエージェント $A_j$ からの同一目標のグローバル座標系の目標状態ベクトルに基づいて更新するという局所的な相互作用により，大域的な機能として，同一目標を媒体としてセンサネットワーク全体のバイアスを推定・補償するものである [4], [5]．

## 3. 自律分散型センサフュージョンのためのバイアス推定

ここでは，図 2 に示す自律分散型センサフュージョンにおいてバイアスを推定・補償するマルチエージェント型バイアス推定法（以降，提案法と呼ぶ）について述べる．

### 3.1 提案法の処理概要

$N$ 個のセンサ $S_i$ から構成されるセンサネットワークにおいて，センサ $S_i$ に対応するエージェント $A_i$ は，センサ $S_i$ の位置を座標原点とするローカル座標系と，全エージェントが共有するグローバル座標系の 2 つの座標系を持ち，その処理概要は，以下のとおりとなる．

センサ $S_i$ に対応するエージェント $A_i$ は，センサ $S_i$ が観測している目標 $T_l$ ($l \in \{1, ..., L\}$) に対する時刻 $t_{k-1}$ のバイアスを考慮したグローバル座標系の目標状態ベクトル $\hat{\mathbf{x}}_{il}(k-1)$ とセンサ $S_j$ に対応するエージェント $A_j$ からの目標 $T_l$ のバイアスを考慮したグローバル座標系の目標状態ベクトル $\hat{\mathbf{x}}_{jl}(k-1)$ を用いて，下式のように，時刻 $t_{k-1}$ の相互作用を考慮したグローバル座標系の目標状態ベクトル（以降，相互作用ベクトルと呼ぶ）$\tilde{\mathbf{x}}_{il}(k-1)$ を算出する．

$$\tilde{\mathbf{x}}_{il}(k-1) = \hat{\mathbf{x}}_{il}(k-1) + \varepsilon_l \sum_j a_{ijl} \left( \hat{\mathbf{x}}_{jl}(k-1) - \hat{\mathbf{x}}_{il}(k-1) \right) \quad (1)$$

ここで，$\varepsilon_l$ は重みを表す．また，$a_{ijl}$ は，センサ $S_i$ とセンサ

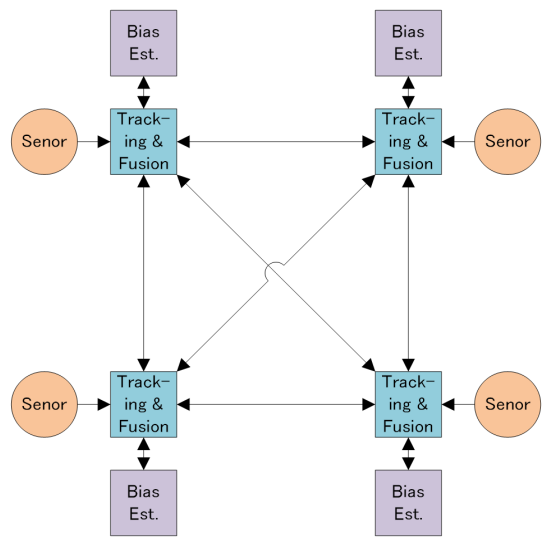

図 2 提案法のシステム構成
Fig. 2 System structure of the proposed method.

$S_j$ が共に目標 $T_l$（すなわち，同一目標）を観測している場合は 1，その他は 0 とする．

次に，センサ $S_i$ に対応するエージェント $A_i$ は，センサ $S_i$ が観測している目標 $T_l$ に対する時刻 $t_{k-1}$ のグローバル座標系の相互作用ベクトル $\tilde{\mathbf{x}}_{il}(k-1)$，同ローカル座標系の目標状態ベクトル $\hat{\mathbf{y}}_{il}(k-1)$，センサ $S_i$ の時刻 $t_{k-1}$ のグローバル座標系のセンサ状態ベクトル $\mathbf{z}_i(k-1)$ を用いて，下式の逆問題を解くことにより，センサ $S_i$ の時刻 $t_{k-1}$ のバイアスベクトル $\mathbf{b}_i(k-1)$ を推定する．

$$\tilde{\mathbf{x}}_{il}(k-1) = f(\hat{\mathbf{y}}_{il}(k-1), \mathbf{z}_i(k-1), \mathbf{b}_i(k-1)) \tag{2}$$

更に，センサ $S_i$ に対応するエージェント $A_i$ は，センサ $S_i$ が観測している目標 $T_l$ に対する時刻 $t_k$ のローカル座標系の目標状態ベクトル $\hat{\mathbf{y}}_{il}(k)$，センサ $S_i$ の時刻 $t_k$ のグローバル座標系のセンサ状態ベクトル $\mathbf{z}_i(k)$，センサ $S_i$ の時刻 $t_{k-1}$ のバイアスベクトル $\mathbf{b}_i(k-1)$ を用いて，下式のように，時刻 $t_k$ のバイアスを考慮したグローバル座標系の目標状態ベクトル $\hat{\mathbf{x}}_{il}(k)$ を算出する．

$$\hat{\mathbf{x}}_{il}(k) = f(\hat{\mathbf{y}}_{il}(k), \mathbf{z}_i(k), \mathbf{b}_i(k-1)) \tag{3}$$

提案法では，上記の処理を繰り返し実施することで，エージェントが算出する同一目標のバイアスを考慮したグローバル座標系の目標状態ベクトルを漸近的に一致させ，センサネットワーク全体のバイアスを推定・補償する．

### 3.2 提案法の処理内容の具体例

次に，距離系と角度系にバイアスが存在する移動しない 2 次元センサから構成されるセンサネットワークを例に，提案法の処理内容の具体例を述べる．

センサ $S_i$ のグローバル座標系のセンサ状態ベクトル $\mathbf{z}_i(k)$ は，下式で表されるものとする．

$$\mathbf{z}_i(k) = \left[ \begin{array}{c} x_i \\ y_i \end{array} \right] \tag{4}$$

ここで，$x_i$ と $y_i$ は，それぞれグローバル座標系における $x$ 軸と $y$ 軸の位置の座標値を表す．

センサ $S_i$ が観測している目標 $T_l$ に対する時刻 $t_k$ の（センサ $S_i$ の位置を座標原点とする）ローカル座標系の目標状態ベクトル $\hat{\mathbf{y}}_{il}(k)$ は，下式で表されるものとする．

$$\hat{\mathbf{y}}_{il}(k) = \left[ \begin{array}{c} r_{il}(k) \\ \theta_{il}(k) \end{array} \right] \tag{5}$$

ここで，$r_{il}(k)$ は，距離系のバイアス（未知数）の影響を受けたセンサ $S_i$ から目標 $T_l$ までの距離を表し，$\theta_{il}(k)$ は，角度系のバイアス（未知数）の影響を受けたセンサ $S_i$ から目標 $T_l$ への角度を表す．

また，センサ $S_i$ の時刻 $t_k$ のバイアスベクトル $\mathbf{b}_i(k)$ は，下式で表されるものとする．

$$\mathbf{b}_i(k) = \left[ \begin{array}{c} \Delta r_i(k) \\ \Delta \theta_i(k) \end{array} \right] \tag{6}$$

ここで，$\Delta r_i(k)$ と $\Delta \theta_i(k)$ は，それぞれ距離系と角度系のバイアスの推定値を表す．

このとき，式 (1) に対応し，センサ $S_i$ の目標 $T_l$ に対する時刻 $t_k$ のグローバル座標系の相互作用ベクトル $\tilde{\mathbf{x}}_{il}(k)$ は，下式で表されるものとする．

$$\tilde{\mathbf{x}}_{il}(k) = \left[ \begin{array}{c} \tilde{x}_{il}(k) \\ \tilde{y}_{il}(k) \end{array} \right] \tag{7}$$

ここで，相互作用ベクトル $\tilde{\mathbf{x}}_{il}(k)$ の要素である $\tilde{x}_{il}(k)$ と $\tilde{y}_{il}(k)$ は，それぞれグローバル座標系の $x$ 軸と $y$ 軸の座標値を表す．

同様に，式 (3) に対応し，センサ $S_i$ の目標 $T_l$ に対する時刻 $t_k$ のバイアスを考慮したグローバル座標系の目標状態ベクトル $\hat{\mathbf{x}}_{il}(k)$ は，下式で表されるものとする．

$$\hat{\mathbf{x}}_{il}(k) = \left[ \begin{array}{c} \hat{x}_{il}(k) \\ \hat{y}_{il}(k) \end{array} \right] \tag{8}$$

ここで，目標状態ベクトル $\hat{\mathbf{x}}_{il}(k)$ の要素である $\hat{x}_{il}(k)$ と $\hat{y}_{il}(k)$ は，それぞれグローバル座標系の $x$ 軸と $y$ 軸の座標値を表す．

また，便宜上，変数の時刻を表すパラメータを省略し，相互作用ベクトル $\tilde{\mathbf{x}}_{il}(k)$ または目標状態ベクトル $\hat{\mathbf{x}}_{il}(k)$ を $\mathbf{x}_{il}$ で表すと，$\mathbf{x}_{il} = f(\hat{\mathbf{y}}_{il}, \mathbf{z}_i, \mathbf{b}_i)$ の変換は，下式となる．

$$\begin{aligned} \left[ \begin{array}{c} x_{il} \\ y_{il} \end{array} \right] &= f \left( \left[ \begin{array}{c} r_{il} \\ \theta_{il} \end{array} \right], \left[ \begin{array}{c} x_i \\ y_i \end{array} \right], \left[ \begin{array}{c} \Delta r_i \\ \Delta \theta_i \end{array} \right] \right) \\ &= \left[ \begin{array}{c} (r_{il} - \Delta r_i) \sin(\theta_{il} - \Delta \theta_i) + x_i \\ (r_{il} - \Delta r_i) \cos(\theta_{il} - \Delta \theta_i) + y_i \end{array} \right] \end{aligned} \tag{9}$$

ここでセンサ数 $N = 4$，目標数 $L = 4$ として，センサ $S_1$ が，センサ $S_2$ と共に目標 $T_1$ を観測し，センサ $S_4$ と共に目標 $T_4$ を観測している状況を例に，センサ $S_1$ に対応するエージェント $A_1$ が行う処理内容を説明する．

エージェント $A_1$ は，グローバル座標系の相互作用ベクトル $\tilde{\mathbf{x}}_{11}(k-1)$ と $\tilde{\mathbf{x}}_{14}(k-1)$ を算出する．グローバル座標系の相互作用ベクトル $\tilde{\mathbf{x}}_{11}(k-1)$ は，センサ $S_1$ の目標 $T_1$ に対するバイアスを考慮したグローバル座標系の目標状態ベクトル $\hat{\mathbf{x}}_{11}(k-1)$ と，共に目標 $T_1$ を観測しているセンサ $S_2$ に対応するエージェント $A_2$ が算出したバイアスを考慮したグローバル座標系の目標状態ベクトル $\hat{\mathbf{x}}_{21}(k-1)$ を用いて，下式で算出する．

$$\begin{aligned} \tilde{\mathbf{x}}_{11}(k-1) &= \hat{\mathbf{x}}_{11}(k-1) \\ &\quad + \varepsilon_1 \sum_j a_{1j1} \left( \hat{\mathbf{x}}_{j1}(k-1) - \hat{\mathbf{x}}_{11}(k-1) \right) \\ &= \hat{\mathbf{x}}_{11}(k-1) + \varepsilon_1 \left( \hat{\mathbf{x}}_{21}(k-1) - \hat{\mathbf{x}}_{11}(k-1) \right) \\ &= (1 - \varepsilon_1) \hat{\mathbf{x}}_{11}(k-1) + \varepsilon_1 \hat{\mathbf{x}}_{21}(k-1) \end{aligned} \tag{10}$$

同様に，グローバル座標系の相互作用ベクトル $\tilde{\mathbf{x}}_{14}(k-1)$ は，センサ $S_1$ の目標 $T_4$ に対するバイアスを考慮したグローバル座標系の目標状態ベクトル $\hat{\mathbf{x}}_{14}(k-1)$ と，共に目標 $T_4$ を観測しているセンサ $S_4$ に対応するエージェント $A_4$ が算出したバイア

スを考慮したグローバル座標系の目標状態ベクトル $\hat{\mathbf{x}}_{44}(k-1)$ を用いて，下式で算出する．

$$\begin{aligned}
\tilde{\mathbf{x}}_{14}(k-1) &= \hat{\mathbf{x}}_{14}(k-1) \\
&\quad + \varepsilon_4 \sum_j a_{1j4}(\hat{\mathbf{x}}_{j4}(k-1) - \hat{\mathbf{x}}_{14}(k-1)) \\
&= \hat{\mathbf{x}}_{14}(k-1) + \varepsilon_4 (\hat{\mathbf{x}}_{44}(k-1) - \hat{\mathbf{x}}_{14}(k-1)) \\
&= (1-\varepsilon_4)\hat{\mathbf{x}}_{14}(k-1) + \varepsilon_4 \hat{\mathbf{x}}_{44}(k-1)
\end{aligned} \quad (11)$$

次に，エージェント $A_1$ は，算出したグローバル座標系の相互作用ベクトル $\tilde{\mathbf{x}}_{11}(k-1)$ と $\tilde{\mathbf{x}}_{14}(k-1)$，ローカル座標系の目標状態ベクトル $\hat{\mathbf{y}}_{11}(k-1)$ と $\hat{\mathbf{y}}_{14}(k-1)$，センサ $S_1$ のグローバル座標系のセンサ状態ベクトル $\mathbf{z}_1(k-1)$ を用いて，下式の逆問題を解くことにより，センサ $S_1$ の時刻 $t_{k-1}$ のバイアスベクトル $\mathbf{b}_1(k-1)$ を推定する．

$$\begin{cases} \tilde{\mathbf{x}}_{11}(k-1) = f(\hat{\mathbf{y}}_{11}(k-1), \mathbf{z}_1(k-1), \mathbf{b}_1(k-1)) \\ \tilde{\mathbf{x}}_{14}(k-1) = f(\hat{\mathbf{y}}_{14}(k-1), \mathbf{z}_1(k-1), \mathbf{b}_1(k-1)) \end{cases} \quad (12)$$

更に，エージェント $A_1$ は，センサ $S_1$ が観測している目標 $T_1$ と $T_4$ に対する時刻 $t_k$ のローカル座標系の目標状態ベクトル $\hat{\mathbf{y}}_{11}(k)$ と $\hat{\mathbf{y}}_{14}(k)$，センサ $S_1$ のグローバル座標系のセンサ状態ベクトル $\mathbf{z}_1(k)$，センサ $S_1$ の時刻 $t_{k-1}$ のバイアスベクトル $\mathbf{b}_1(k-1)$ を用いて，センサ $S_1$ のバイアスを考慮したグローバル座標系の目標状態ベクトル $\hat{\mathbf{x}}_{11}(k)$ と $\hat{\mathbf{x}}_{14}(k)$ を算出する．

$$\begin{cases} \hat{\mathbf{x}}_{11}(k) = f(\hat{\mathbf{y}}_{11}(k), \mathbf{z}_1(k), \mathbf{b}_1(k-1)) \\ \hat{\mathbf{x}}_{14}(k) = f(\hat{\mathbf{y}}_{14}(k), \mathbf{z}_1(k), \mathbf{b}_1(k-1)) \end{cases} \quad (13)$$

そして，それぞれのセンサに対応するエージェントが，上記と同様の処理を繰り返し実施する．

## 4. シミュレーション

距離系と角度系にバイアスが存在する2次元センサから構成されるセンサネットワークを用いた数値シミュレーションにより，提案法の基本機能を確認する．

### 4.1 シミュレーション条件

シミュレーション条件1では，図3に示すように，2次元空間のそれぞれ (0km, 0km) と (10km, 0km) に固定センサ $S_1$ と $S_2$ が配置され，それぞれ (0km, 5km) と (10km, 5km) の位置にある固定目標 $T_1$ と $T_2$ に対して，センサ $S_1$ と $S_2$ が共に目標 $T_1$ と $T_2$ を観測しているものとする．また，シミュレーション条件2では，図4に示すように，2次元空間のそれぞれ (0km, 0km) と (10km, 0km) に固定センサ $S_1$ と $S_2$ が配置され，それぞれ (0km, 10km) と (10km, 10km) の位置にある固定目標 $T_1$ と $T_2$ に対して，センサ $S_1$ と $S_2$ が共に目標 $T_1$ と $T_2$ を観測しているものとする．ここで，センサ $S_1$ のバイアスの真値は距離系 10m，角度系 $0.1°$ とし，センサ $S_2$ のバイアスの真値は距離系 20m，角度系 $0.2°$ とする．また，各センサの観測誤

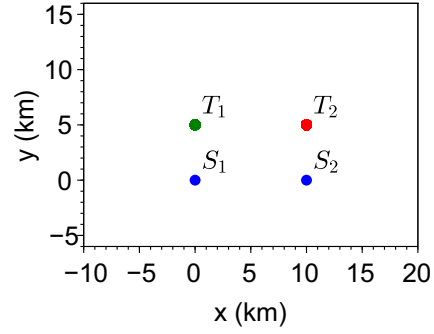

図 3　シミュレーション条件1（センサと目標の配置）
Fig. 3　Simulation scenario 1.

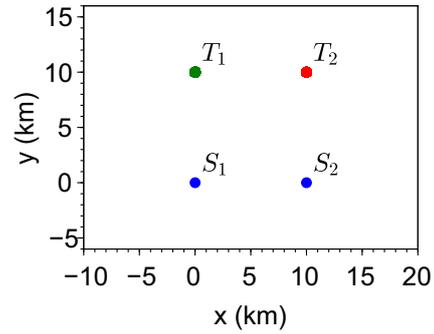

図 4　シミュレーション条件2（センサと目標の配置）
Fig. 4　Simulation scenario 2.

差は，平均 0 の白色正規雑音とし，標準偏差を距離系 10m，角度系 $0.1°$ とする．

### 4.2 シミュレーション結果

20回のモンテカルロシミュレーションを実施し，RMS（Root Mean Square）を評価する．

シミュレーション条件1に対応するシミュレーション結果として，グローバル座標系の目標状態ベクトル間の位置の差を図5，距離系と角度系のバイアス推定値の残差（バイアス推定値とバイアス真値の差）を図6に示す．同様に，シミュレーション条件2に対応するシミュレーション結果を図7と図8に示す．

図5と図7のシミュレーション結果から，バイアスを考慮したグローバル座標系の目標状態ベクトル間の位置の差（図中の「w BC」）は，バイアスを考慮しないグローバル座標系の目標状態ベクトル間の位置の差（図中の「w/o BC」）よりも，グローバル座標系の目標状態ベクトル間の位置の差が小さくなっており，提案法のバイアス推定・補償の効果を確認できる．

また，図6のシミュレーション結果から，横軸の時間 $t_k$ が増加するにつれてバイアス推定値の残差が 0 に漸近しており，バイアス推定値がバイアス真値に近づいている（すなわち，バイアス推定・補正ができている）ことが分かる一方，図8（特に距離系の残差）からは，バイアス推定値がバイアス真値に近づいておらず，バイアス補正ではなくバイアス補償となっていることが分かる．

## 5. まとめ

本稿では，自律分散型センサフュージョンにおけるバイアスの推定・補償を目的とし，センサネットワーク全体を複数エージェントの局所的な相互作用をもとに大域的な機能を発現するマルチエージェントシステムとして捉え，マルチエージェントモデルを用いてセンサのバイアスを推定・補償する新しいバイアス推定法を提案した．

簡易なモデルを用いた計算機シミュレーションにより基本機能の確認を行い，センサのバイアスを推定・補償できることを確認した．

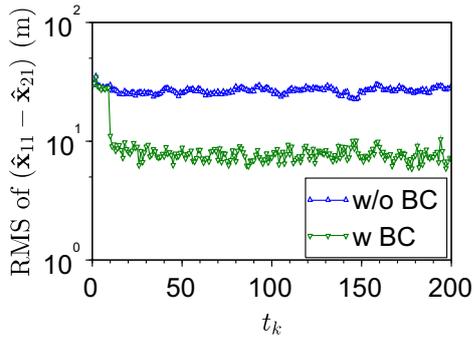
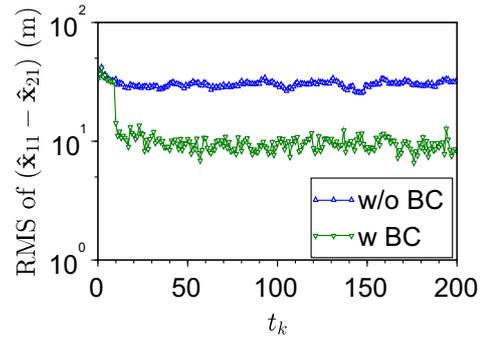
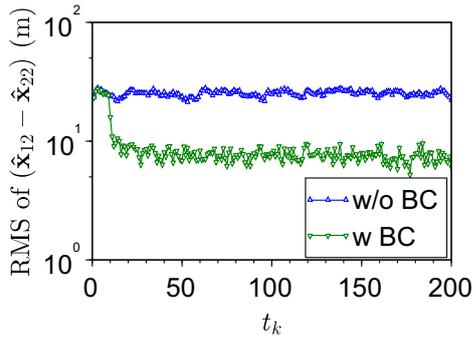
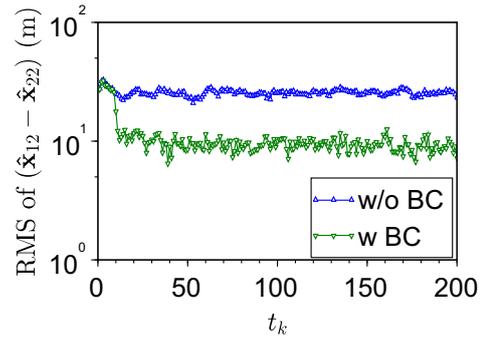

図 5  2つのグローバル座標系の目標状態ベクトル間の位置の差
（シミュレーション条件 1）
Fig. 5  Difference between the two tracks (simulation scenario 1).

図 7  2つのグローバル座標系の目標状態ベクトル間の位置の差
（シミュレーション条件 2）
Fig. 7  Difference between the two tracks (simulation scenario 2).

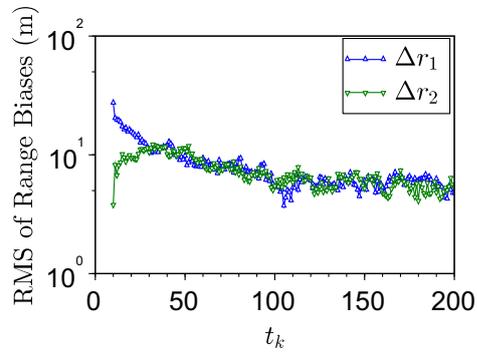
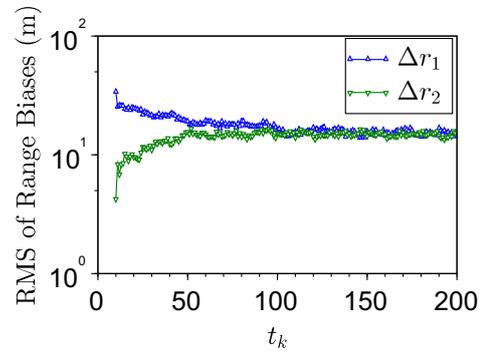

図 6  距離系と角度系のバイアス推定値の残差
（シミュレーション条件 1）
Fig. 6  Residual error in estimated bias of range and angle (simulation scenario 1).

図 8  距離系と角度系のバイアス推定値の残差
（シミュレーション条件 2）
Fig. 8  Residual error in estimated bias of range and angle (simulation scenario 2).